\documentclass[aps,pre,twocolumn,groupedaddress,showpacs,amssymb,amsmath]{revtex4-1}
\usepackage{lipsum}
\usepackage{graphicx}   
\usepackage{verbatim}   
\usepackage{color}      
\usepackage{subfigure}  
\usepackage{hyperref} 
\usepackage{amsmath} 
\usepackage{color}   
\newcommand{\be}{\begin{equation}}
\newcommand{\ee}{\end{equation}}
\begin{document}
\title{Exchange
fluctuation theorem for heat transport between multi-terminal harmonic systems }


\author{Bijay Kumar Agarwalla$^{1,2}$}
\author{Huanan Li$^{1}$}
\author{Baowen Li$^{1,3,4}$}
\author{Jian-Sheng Wang$^{1}$}

\affiliation{$^{1}$ Department of Physics and Center for Computational Science and Engineering,
National University of Singapore, Singapore 117542, Republic of Singapore\\
$^{2}$ Chemistry Department, University of California, Irvine, California 92697-2025, USA \\
$^{3}$ NUS Graduate School for Integrative Sciences and Engineering, Singapore 117456, Republic of Singapore\\
$^{4}$ Center for Phononics and Thermal Energy Science, Department of Physics, Tongji University, 200092 Shanghai, China }



\date{\today}
\begin{abstract}
We study full counting statistics for transferred heat and
entropy production between multi-terminal systems in absence of a finite junction. The systems are modelled as collections of coupled
harmonic oscillators which are kept at different equilibrium temperatures and
are connected via arbitrary time dependent couplings. Following consistent
quantum framework and two-time measurement concept we obtain analytical
expressions for the generalized cumulant generating function. We discuss transient and steady-state fluctuation theorems for the transferred quantities. 
We also address the effect of coupling strength on the
exchange fluctuation theorem.

\end{abstract}
\maketitle

\section{Introduction}

\begin{figure}
\includegraphics[width=0.65\columnwidth]{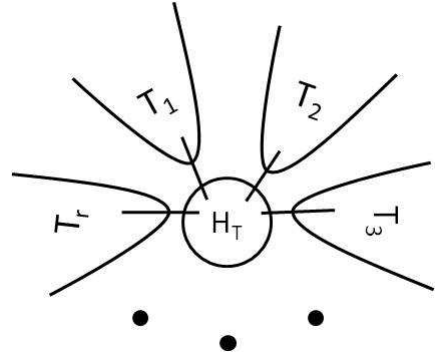}
\caption{\label{set-up} A schematic representation for exchange fluctuation theorem setup consists of multi-terminals without a junction. The terminals are at their respective equilibrium temperatures $T_{\alpha}=(k_{B}\beta_{\alpha})^{-1}$. The reservoirs are interacting via the Hamiltonian ${\cal H}_{T}(t)$ which is switched on at $t=0$.}
\end{figure}

A primary interest in the field of nonequilibrium statistical physics is to understand the
transport properties of energy and matter. An important class of experimental
devices for such study consists of several reservoirs, maintained at different temperatures
and/or chemical potentials and interacting via time-dependent or static coupling,
see fig.~(\ref{set-up}). Using this type of setup numerous analyses have been carried 
out to investigate what happens to particle/energy current from the perspective of modelling 
thermal switch, heat pump, nano-oscillators, etc
\cite{Eduardo,Eduardo-electron}. 

However, in addition to the particle/energy current, the stochastic nature of the
reservoirs generates quantum and thermal fluctuations to these quantities which are not
arbitrary but rather related via universal fluctuation relations
\cite{Hanggi-review, Esposito-review, Jarzynski-review,fluc-theorems}. For
the setup in fig.~(\ref{set-up}) fluctuation relations have been referred to as ``exchange fluctuation
theorems'' (XFT). Using principle of micro-reversibility of the underlying
Hamiltonian dynamics, Jarzynski and W\'ojcik \cite{Jarzynski} put forward the
first quantum XFT which states that $\langle e^{-\Delta \beta Q_{L}}
\rangle_{t_M} =1$ between two weakly connected system $L$ and $R$. Here $\Delta
\beta=\beta_R -\beta_L$, $\big(\beta_{\alpha}=1/(k_{\rm B}T_{\alpha})\big)$ is the thermodynamic 
affinity and $Q_L$ is
the amount of heat transferred out of the left reservoir over the time interval
$[0,t_M]$. A more general version of this XFT for multi-terminal system was
later derived by Andrieux et al \cite{Andrieux} which states that $\langle
e^{-\Sigma}\rangle_{t_M}=1$, where $\Sigma=-\sum_{\alpha=1}^{r}\beta_\alpha
Q_\alpha$ is the net entropy production due to exchange of heat. Here $r$ is the number of reservoirs.
This relation is valid for arbitrary time-dependent coupling strength between the
reservoirs and reduces to Jarzynski and W\'ojcik relation for $r=2$ in the limit
of weak coupling. To get a deeper insight of these relations we choose a particular model system and study the complete statistics which is also known as ``full counting statistics'' (FCS) \cite{Belzig,
Nazarov,Schonhammer} in the electronic and photonic literature. The first result for FCS
study was obtained for non-interacting electrons by Levitov and Lesovik
\cite{Levitov} using scattering picture. In the phononic case Saito and Dhar \cite{Saito-Dhar} and later
Wang et al.\ \cite{fcs-bijay} found an explicit expression for the cumulant
generating function (CGF) of heat in presence of a finite junction. Master equation approach has also been developed to study FCS for markovian \cite{Renjie} as well as non-Markovian systems \cite{Flindt}. However,
the study of CGF for a multi-terminal setup in the absence of a junction such as in Fig.~1 in the context of XFT has not been addressed
yet. Very recently experimental verification of XFT is reported for
electrons \cite{Utsumi}. For phonons nano-resonator seems to be a potential candidate for
performing such experiments \cite{measurement,AAClerk-2011}.
  
In this paper, we focus on a multi-terminal system {\textit {without}} a finite junction
part which are interacting via arbitrary time-dependent coupling. We consider that the systems are collection of finite number of
coupled harmonic oscillators. For infinite oscillators these systems behave as phononic heat baths and often go by the name of Rubin \cite{Rubin}. The interaction between the systems are chosen to be bilinear in position with time-dependent coupling. An analogue of this model for electrons is the
well-known tight binding Hamiltonian. Since harmonic systems never equilibrates, here we
investigate both the transient and steady-state behavior and also the
corresponding fluctuation theorems for heat and entropy production. For two-terminal case ($r=2$) we also discuss the effect of coupling strength
on exchange fluctuation theorem (XFT). 

The paper is organized as follows. We start in Sec.~II by introducing our model. Then in Sec.~III we discuss the initial preparations of
the system. In Sec.~IV  we obtain the analytical expressions for the generalized CGF using non-equilibrium Green's function method.
Then in Sec.~V we obtain the long-time limit of the CGF. In Sec.~VI we discuss our numerical findings for heat exchange for
two-terminal situation and discuss about XFT for different coupling strengths and finite size effects of the systems on the cumulants of heat. Finally we conclude with a short discussion in Sec.~VIII.

\section{Model Hamiltonian}
We consider arbitrary number of systems, say $r$, each consists of finite/infinite number of coupled harmonic oscillators and is given by the Hamiltonian
\begin{equation}
{\cal H}_{\alpha}= \sum_{i=1}^{N_{\alpha}} \frac{(p^{\alpha}_{i})^{2}}{2} + \sum_{i,j=1}^{N_{\alpha}} \frac{1}{2}\, K_{ij}^{\alpha} u_{i}^{\alpha} u_{j}^{\alpha}, \quad \alpha=1,2, \cdots r,
\end{equation}
where $p^{\alpha}_{i}$ is the momentum of the $i$-th particle in the $\alpha$-th
reservoir, $u^{\alpha}_{i} (= \sqrt{m_i}\, x_{i}^{\alpha})$ is the mass normalized position operators. These operators obey
the usual Heisenberg commutation relations $\left[ u_j^{\alpha}(t), {p}_k^{\beta}(t)
\right] = i \hbar\,\delta_{jk}
  \,\delta^{{\alpha}{\beta}},~\alpha, \beta = {1,2,\cdots, r}$. $N_{\alpha}$ is
the number of degrees of freedom in each system, $K^{\alpha}$ is the force constant
matrix. 

The interacting Hamiltonian ${\cal H}_{T}(t)$ between the systems is taken in
bilinear form and couples only the position operators of the systems via
arbitrary time-dependent coupling matrix $V^{\alpha \beta}(t)$ and is written in
the following form
\begin{equation}
{\cal H}_{T}(t)=\frac{1}{2} \sum_{\alpha \neq \beta} u_{\alpha}^{T} V^{\alpha
\beta}(t) u_{\beta}.
\end{equation}
We assume that the interaction between the systems is switched on from $t=0$. Therefore the total Hamiltonian after the connection is 
\begin{equation}
{\cal H}(t)= \sum_{\alpha=1}^{r} {\cal H}_{\alpha} + {\cal H}_{T}(t).
\end{equation}

In this paper, we will present numerical results for the cumulants of exchanged heat between two-terminals ($L$ and $R$) for two specific forms of the coupling: one is $V^{LR}(t)=\theta(t)\,V^{LR}$, where $\theta(t)$ is the Heaviside step function corresponding to the sudden connection between the two systems; the other form of the coupling we choose is $V^{LR}(t)=V^{LR}\,\tanh(\omega_d \,t)$, where $\omega_d$ is the driving frequency. This particular form of the coupling mimics the gradual switch on of the interaction i.e., from adiabatic $(\omega_d \to 0)$ to sudden $(\omega_d \to \infty)$, between the systems.  
Other forms of time-dependent coupling can also be handled easily in this formalism.

\section{Initial Preparation of the systems}
Here we consider the following approach for the initial preparation of the systems. We first allow the systems $\alpha=1,2, \cdots r$ to equilibrate at respective temperatures $T_{\alpha}$ by keeping them in weak contact with the thermal baths. So, the initial density matrix at $t=0$ is of decoupled form with each system at their respective equilibrium distribution (canonical) 
\begin{equation}
\rho_0=\prod_{\alpha=1}^{r} \frac{e^{-\beta_{\alpha} {\cal H}_{\alpha}}} {{\rm Tr} (e^{-\beta_{\alpha} {\cal H}_{\alpha}})}.
\label{density-operator}
\end{equation}
Then we disconnect the systems from their respective heat baths and switch-on the interaction at $t=0$ to allow the exchange of heat up to a maximum time $t_M$ and then switch off the interaction suddenly. Because of such initial preparation of the systems the net heat transferred is not given by a single number but rather a nontrivial probability distribution. 

In a more general scenario we can talk about the joint probability distribution $P(\{Q_{\alpha}\})$ where $\{Q_{\alpha}\}$ is the shorthand notation for heat exchanged between all the systems $Q_{1}, Q_{2}, \cdots, Q_{r}$. Such a general distribution contains valuable informations such as multiple-point correlations between heat say $\langle Q_{1}(t_M) Q_2(t_M)\rangle$ or $\langle Q_{1}(t_M) Q_2(t_M) Q_{r}(t_M)\rangle$ etc. To obtain $P(\{Q_{\alpha}\})$ it is convenient to look at the characteristic function (CF) corresponding to this distribution. 

In the following we first define heat and entropy production and the corresponding CF's and then obtain analytical expressions according to the Keldysh nonequilibrium Green's function technique.

\section{Definition of heat, entropy production and corresponding characteristic functions}
We are interested in the statistics of exchanged heat (integrated current) and entropy production in a given time interval $[0,t_M]$. Based on the observation that heat is not a quantum observable \cite{Hanggi-review}, we define heat flowing out of a system (say $\alpha$), as the difference of the system Hamiltonian at time $t=0$ and $t_M$ i.e.,
\begin{equation}
{\cal Q}_{\alpha}(t_M)\equiv {\cal H}_{\alpha}(0)\!-\!{\cal H}_{\alpha}^{H}(t_M),\,\,\alpha=1,2, \cdots r,
\end{equation}
where ${\cal H}_{\alpha}^{H}(t)$ is the Hamiltonian in the Heisenberg picture and evolving with respect to the full Hamiltonian ${\cal H}(t)$ i.e., ${\cal H}_{\alpha}^{H}(t)={\cal U}^{\dagger}(t,0)\,{\cal H}_{\alpha}\,{\cal U}(t,0)$ and ${\cal U}(t,t')=T \exp \big[{-\frac{i}{\hbar} \int_{t'}^{t} dt_1 {\cal H}(t_1)}\big]$ for $t>t'$. Here $T$ is the time-ordered operator which orders operators from left to right with decreasing time argument. Based on the definition of heat we also define the net entropy production for this nonequilibrium process as
\begin{equation}
\Sigma(t_M)=-\sum_{\alpha=1}^{r} \beta_{\alpha} {\cal Q}_{\alpha}(t_M).
\label{sigma}
\end{equation}
In order to obtain the CF for entropy production, we have to introduce counting parameter $\chi_{\alpha}$ for each $Q_{\alpha}$. 
To construct the CF we follow two-time quantum measurement method \cite{Neumann, Esposito-review} i.e., we perform measurements for all system Hamiltonians ${\cal H}_{\alpha}, \alpha=1,2, \cdots r$ at $t=0$ and $t=t_M$. Let at $t=0$ and $t=t_M$ the outcomes of the measurements are set of eigenvalues $\{\epsilon_{m}^{\alpha}\}$ and $\{\epsilon_{n}^{\alpha}\}$ respectively where $\{\epsilon_{m}^{\alpha}\}= (\epsilon_{m}^{1}, \epsilon_{m}^{2} \cdots \epsilon_{m}^{r})$ and similarly for $\{\epsilon_{n}^{\alpha}\}$. So for each realization exchanged heat is given by $Q_{\alpha}= \epsilon_{m}^{\alpha} - \epsilon_{n}^{\alpha}, \forall\, \alpha= 1, 2, \cdots r$. Now summing over all such possible realizations we can construct the generalized CF as \cite{fluc-theorems}
\begin{equation} 
{\cal Z}({\vec{\chi}})= \Big \langle W^{\dagger} \, {\cal U}^{\dagger}(t_M,0)\, W^{2}\, {\cal U}(t_M,0)\, W^{\dagger} \Big \rangle,
\end{equation}
where we define the vector ${\vec{\chi}}=(\chi_1, \chi_2, \cdots \chi_r)$, the set of counting fields and $W=\prod_{\alpha=1}^{r} W_{\alpha}= \prod_{\alpha=1}^{r} \exp(-i \chi_{\alpha} {\cal H}_{\alpha}/2)=\exp(-i {\vec{\chi}}.{\vec{\cal H}}_{0}/2)$ where ${\vec{\cal H}}_{0}=({\cal H}_1, {\cal H}_{2} \cdots {\cal H}_{r})$.
The average here is with respect to the initial density matrix $\rho_0$ given in Eq.~(\ref{density-operator}). Note that the above expression for CF is also valid for interacting systems.




In order to obtain explicit expression for ${\cal Z}({\vec{\chi}})$ the main objective is to write it
as an effective evolution of the unitary operators with respect to the counting parameters. 
To achieve this one can simplify the above expression and express the CF as 
\begin{equation}
{\cal Z}({\vec{\chi}})=\Big\langle {\cal U}_{{\vec{\chi}/2}}^{\dagger}(t_M,0)\, {\cal U}_{{-\vec{\chi}/2}}(t_M,0) \Big\rangle ,
\label{evolution}
\end{equation}
where ${\cal U}_{\vec{X}}(t,0)=T \exp\big[{-\frac{i}{\hbar} \int_{0}^{t} dt' {\cal H}_{\vec{X}}(t')}\big]$, $(\vec{X}\!=\!\pm \vec{\chi}/2)$ and 
\begin{align}
 {\cal H}_{\vec{X}}(t)&=e^{i \vec{X}\cdot {\vec{\cal H}}_{0}} {\cal H}(t) e^{-i \vec{X}\cdot {\vec{\cal H}}_{0}} \nonumber \\
&=\sum_{\alpha=1}^{r} {\cal H}_{\alpha} + \frac{1}{2} \sum_{\alpha \neq \beta} u_{\alpha}^{T}(\hbar X_{\alpha}) V^{\alpha \beta}(t) u_{\beta}(\hbar X_{\beta}).
\end{align}
It is clear from the last expression that the effect of the measurement generates a time translation to the position operators of the interaction Hamiltonian $W_{\alpha}^{\dagger} u_{\alpha} W_{\alpha} \rightarrow u_{\alpha}(\hbar \chi_{\alpha}/2)$. In a similar way it can be shown that for electrons the measurement of the number operator (electron counting) generates a phase in the interaction Hamiltonian \cite{fcs-bijay}.

Now transforming to the interaction picture with respect to the decoupled Hamiltonian $h=\sum_{\alpha=1}^{r} {\cal H}_{\alpha}$, the above expression on the Keldysh contour $[0,t_M]$ reads as 
\begin{equation}
{\cal Z}(\vec{\chi})=\Big \langle T_c \exp\Big[{-\frac{i}{\hbar} \int_c d\tau {\cal H}_{X(\tau)}^{T,I}} \Big]\Big \rangle.
\label{interaction-Z}
\end{equation}
\begin{figure}
\includegraphics[width=0.55\columnwidth]{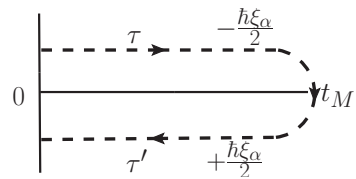}
\caption{The complex-time Keldysh contour for FCS problem. The path of the contour begins at time $t=0$ , goes to time $t_M$ with counting field $-\frac{\hbar \xi_{\alpha}}{2}$, and then goes
back to time $t = 0$ with counting field $+\frac{\hbar \xi_{\alpha}}{2}$ . Here $\tau$ and $\tau'$ are complex-time variables along the contour.
}
\end{figure}
Here $T_c$ is the contour-ordered operator which orders
operators according to their contour time argument, earlier
contour time places an operator to the right. The tunneling Hamiltonian on the contour (in the interaction picture) is expressed as
\begin{equation}
{\cal H}_{X(\tau)}^{T,I}= \frac{1}{2} \sum_{\alpha \neq \beta} u_{\alpha}^{T}(\tau+\hbar X_{\alpha}(\tau))\, V^{\alpha \beta}(\tau) \, u_{\beta}(\tau+\hbar X_{\beta}(\tau)),
\end{equation}
where $X_{\alpha}(\tau),X_{\beta}(\tau)$ are now contour time $(\tau)$ dependent functions and take two possible values depending on the location of $\tau$ on the branch i.e.,
\begin{equation}
X_{\alpha}^{\pm}(t)=
\begin{cases}
\mp\frac{\hbar \chi_{\alpha}}{2} \quad \text{for} \quad 0 \leq t\leq t_M\\
 0 \,\,\,\, \text{for} \quad t<0 \,\,\,\, \text{and}\,\,\,\, t>t_M.
\end{cases}
\end{equation}
The plus and minus sign in the superscript of $X_{\alpha}$ corresponds to the upper and lower branch of the contour respectively (see Fig.~2).

We now consider the general properties of the cumulant generating function (CGF) ${\cal F}(\vec{\chi})$ defined as the logarithm of the CF i.e., ${\cal F}(\vec{\chi})\equiv\ln {\cal Z}(\vec{\chi})$. Expanding the exponential in Eq.~(\ref{interaction-Z}) and using Feynman diagrammatic technique \cite{recent-review} where all diagrams are closed ring type and finally making use of the linked-cluster theorem (sum of all connected diagrams) the infinite series can be summed up exactly and can be written as
\begin{eqnarray}
{\cal F}(\vec{\chi})&=&-\frac{1}{2} \ln \det \Big(\bf{I}- \bf{V} {\bf g}^{X}\Big) \nonumber \\
&=&-\frac{1}{2} {\rm Tr}_{j,\tau} \ln \Big(\bf{I}- \bf{V} {\bf {g}}^{X}\Big),
\label{CGF-1}
\end{eqnarray}
(Bold symbol refers to matrix in the discretized contour time) where ${\bf{V}}$ is a $r \times r$ off-diagonal matrix with matrix elements $V_{\alpha \beta}(\tau,\tau')=V_{\alpha \beta}(\tau) \, \delta (\tau,\tau')$ with $\alpha,\beta=1,2, \cdots r$ 
and $\bf{{g}}^{X}$ is a $r \times r$ diagonal matrix with matrix elements ${g}^{X}_{\alpha}(\tau,\tau')$, given as 
\begin{equation}
{g}^{X}_{\alpha}(\tau,\tau')=-\frac{i}{\hbar} \Big\langle T_c u_{\alpha}\big(\tau+\hbar X_{\alpha}(\tau)\big) u_{\alpha}^{T}\big(\tau'+\hbar X_{\alpha}(\tau')\big) \Big \rangle. 
\end{equation}
Here the meaning of trace is in terms of contour variable $\tau$ and space index $j$ and ${\rm Tr}_{j,\tau}[{\bf A} \, {\bf B}]=\int_c d\tau_1 \int_c d\tau_2\, A(\tau_1,\tau_2)\, B(\tau_2,\tau_1)$. Note that only $\bf{{g}^{X}}$ in the above expression is a function of the counting fields $\vec{\chi}$. 
This CGF can be simplified further to explicitly satisfy the normalization condition ${\cal F}(\vec{0})=0$ and can be written as (see appendix A)
\begin{equation}
{\cal F}(\vec{\chi})=-\frac{1}{2} \ln \det \Big(\bf{I}-(\bf{I}+\bf{VG}) \bf{V} {\bf g}^{A} \Big).
\label{CGF-2}
\end{equation}

This is the first central result of our paper. This expression is in general valid for both transient and stationary state, for arbitrary time-dependent couplings between the leads and also valid for finite size of the systems. Also the dimensionality of the systems could be arbitrary. The matrix $\bf{G}$ consists of elements $G_{\alpha \beta}(\tau,\tau')$ and satisfies the following Dyson equation on the Keldysh contour
\begin{equation}
{G}(\tau,\tau')=g(\tau,\tau')+\int_c d\tau_1 \int_c d\tau_2\, g(\tau,\tau_1)\,{V}(\tau_1,\tau_2)\,{G}(\tau_2,\tau').
\label{Dyson-1}
\end{equation}
We define $g^{A}(\tau,\tau')={g}^{X}(\tau,\tau')-g(\tau,\tau')$ which is also a diagonal matrix. It is now easy to see that for $\vec{\chi}=0$, $g^{A}=0$ and therefore the normalization condition ${\cal Z}(\vec{0})=1$ is trivially satisfied. Also note that due to the presence of the coupling matrix $V$ only the surface Green's functions \cite{recent-review} are required to compute. This reduces the numerical complexity of the problem. The explicit form of these different matrices for two-terminal case is discussed in Sec.~VI. For arbitrary measurement time $t_M$ and for $\chi_{\alpha}=-i\beta_{\alpha}\, \forall\, \alpha=1,2, \cdots r$, we show in appendix (see appendix B) that $\langle e^{-\Sigma} \rangle_{t_M}=1$.


From this generalized CGF, the CGF for heat $Q_{\alpha}$ can be obtained trivially by substituting all counting parameters except $\chi_{\alpha}$ to be zero. The cumulants can then be computed by taking derivative of the CGF with respect to the counting field. For example the first and second cumulants are written as
\begin{eqnarray}
\langle \langle Q_{\alpha} \rangle \rangle &=& \frac{\partial {\cal F}(\vec{0})}{\partial (i\chi_{\alpha})}, \nonumber \\
\langle \langle Q_{\alpha} Q_{\beta} \rangle \rangle &=& \frac{\partial^{2} {\cal F}(\vec{0})}{\partial (i\chi_{\alpha}) \partial (i \chi_{\beta})}.
\end{eqnarray}

\section{Long-time result for the CGF for heat and entropy-production}
In this section, we specialize the case where only the lead $\alpha$ is measured and derive the stationary state expression for the CGF of heat $Q_{\alpha}$.
In order to achieve the stationary state with infinite recurrence time we require the following criterions to be satisfied,
\begin{itemize}
\item{The size of all the systems should be infinite, i.e., $N_{\alpha} \rightarrow \infty,\, \alpha=1,2, \cdots, r$ and hence are called dissipative leads, so that the sound waves can't scatter back from the boundaries}.
\item{The final measurement time $t_M$ should approach infinity}.
\item{The coupling matrix $V^{\alpha \beta}(t)$ between the leads should be time-independent}.
\end{itemize}
To derive the long-time limit expression we follow the derivation in Ref.~\onlinecite{Huanan1}. 
Here we briefly outline few important relations. Let us calculate the heat transferred from the $\alpha$-th lead. Using the matrix form for $\bf{V}$ and $\bf{G}$ Eq.~(\ref{CGF-2}) in the frequency domain is written as 
\begin{equation}
{\cal F}(\chi_{\alpha})= -t_M \int_{-\infty}^{\infty} \frac{d\omega}{4\pi} \ln \det
\Big[\bf{I}-\big(\breve{g}_{\alpha}^{-1} \breve{G}_{\alpha \alpha} - \bf{I} \big)
\breve{g}_{\alpha}^{-1} \breve{g}_{\alpha}^{A} \Big],
\end{equation}
where,
\begin{gather}
\breve{g}_{\alpha}=\begin{bmatrix}g_{\alpha}^{r} & g_{\alpha}^{K}\\
0 & g_{\alpha}^{a}
\end{bmatrix},\:\breve{G}_{\alpha \alpha}=\begin{bmatrix}G_{\alpha \alpha}^{r} & G_{\alpha \alpha}^{K}\\
0 & G_{\alpha \alpha}^{a}
\end{bmatrix},\\
\breve{g}_{\alpha}^{A}=\frac{1}{2}\begin{bmatrix}a-b & a+b\\
-a-b & -a+b
\end{bmatrix},\\
a= g_{\alpha}^{>}\left(e^{-i\chi_{\alpha}\hbar\omega}-1\right),\: b= g_{\alpha}^{<}\left(e^{i\chi_{\alpha}\hbar\omega}-1\right).\nonumber
\end{gather}
In order to obtain this we take the following steps. First we write down an explicit
expression for ${\cal F}(\chi_{\alpha})$ in contour-time. Then transform the equation to
real time using Langreth theorem \cite{recent-review}, then perform Keldysh rotation \cite{fcs-bijay} to work
with retarded, advanced and Keldysh Green's functions. Finally invoking time-translational
invariance of all the Green's functions in the steady state i.e., $G(t,t')=G(t-t')$ we perform the Fourier
transform and obtain the above equation. For details of the calculation see Sec.~IV of
Ref.~\onlinecite{Huanan1}.



An important quantity to define in this case is the spectral function $\tilde{\Gamma}_{\alpha}$ given by
\begin{equation}
\tilde{\Gamma}_{\alpha}[\omega]= i \Big[\big(g_{\alpha}^{a}\big)^{-1}[\omega]-\big(g_{\alpha}^{r}\big)^{-1}[\omega]\Big].
\end{equation}
Note that this definition differs from the usual definition of the spectral function in presence of a finite junction which is given as the differece of retarded and advanced self-energies. Now for any isolated system $g_{\alpha}^{r}[\omega]= \big[(\omega+i\eta)^{2}-K^{\alpha}\big]^{-1}$ and $g_{\alpha}^{a}[\omega]=\big[g_{\alpha}^{r}[\omega]\big]^{\dagger}$ where $\eta$ is an infinitesimal positive number used to satisfy the causality condition for the retarded and advanced Green's functions in the time domain i.e., 
$g_{\alpha}^{r}(t)=0$ for $t<0$ and $g_{\alpha}^{a}(t)=0$ for $t>0$. For finite system size $\tilde{\Gamma}_{\alpha}[\omega]=4 \omega \eta$
and in the limit $\eta \to 0^+$ it is zero. However for infinite system size ($K^{\alpha}$ is infinite dimensional matrix) this is no longer true. 

Now in order to get final expression for the steady-state CGF we need to know the different components
of the matrix $\breve{G}_{\alpha \alpha}$.
Two important relations that are required to derive the CGF are
\begin{eqnarray}
G_{\alpha \alpha}^{<}[\omega]&=&-i \sum_{\gamma=1}^{r} f_{\gamma}[\omega] G_{\alpha
\gamma}^{r}[\omega] \tilde{\Gamma}_{\gamma}[\omega] G_{\gamma \alpha}^{a}[\omega]
\nonumber \\
G_{\alpha \alpha}^{r}[\omega]-G_{\alpha \alpha}^{a}[\omega]&=&-i \sum_{\gamma=1}^{r}
G_{\alpha \gamma}^{r}[\omega] \tilde{\Gamma}_{\gamma}[\omega] G_{\gamma
\alpha}^{a}[\omega],
\end{eqnarray}
which are obtained by simplifying the Dyson equation given in Eq.~(\ref{Dyson-1}) in the frequency domain. When all the systems are in thermal equilibrium these equations are related by fluctuation-dissipation theorem.

Substituting the expression for $G_{\alpha \alpha}$ and after a lengthy but straightforward calculation leads to the long-time limit expression for the CGF for heat transfer 
\begin{equation}
\frac{{\cal F}(\chi_{\alpha})}{t_M}=-\int_{-\infty}^{\infty} \frac{d\omega}{4\pi} \ln \det
\Big[I- \sum_{\gamma \neq \alpha=1}^{r} {\cal T}_{\alpha \gamma}[\omega] \,{\cal
K}_{\gamma \alpha}(\chi_{\alpha})\Big],
\label{first-central}
\end{equation}
where the function ${\cal K}_{\gamma \alpha}(\chi_{\alpha})$ is given as
\begin{equation}
{\cal K}_{\gamma \alpha}(\chi_{\alpha})=f_{\alpha}(1+f_{\gamma}) (e^{i\chi_{\alpha} \hbar
\omega}-1)+ f_{\gamma}(1+f_{\alpha}) (e^{-i\chi_{\alpha} \hbar \omega}-1).
\end{equation}
This particular function satisfies the Gallavoti-Cohen symmetry ${\cal K}_{\gamma \alpha}(\chi_{\alpha})={\cal K}_{\gamma \alpha}(-\chi_{\alpha} + i (\beta_{\gamma}-\beta_{\alpha}))$ and ${\cal T}_{\alpha \gamma} [\omega]$ is the transmission matrix between the $\alpha$ and the $\gamma$-th terminal and is of the following form
\begin{equation}
{\cal T}_{\alpha \gamma}[\omega]= G_{\alpha \gamma}^{r} {\tilde \Gamma_{\gamma}} G_{\gamma \alpha}^{a} {\tilde \Gamma_{\alpha}}.
\end{equation}
This expression for transmission function was obtained by Lifa et.~al. \cite{Lifa} for two terminal case and was used to study the interface effects.

So in general for multi-terminal situation the CGF of heat will not follow Gallavoti-Cohen fluctuation symmetry due to the presence of the summation in eq.~(\ref{first-central}). However it can recovered trivially for two-terminal case. 
This CGF is the second central result of our paper. 





The cumulants of heat can now be obtained very easily, for example, the first and second cumulants are given as
\begin{eqnarray}
\frac{\langle \langle Q_{\alpha}\rangle \rangle}{t_M}&=& 
\int_{-\infty}^{\infty} \frac{d\omega}{4\pi}\,\hbar \omega\! \!\sum_{\gamma \neq \alpha =1}^{r} {\rm Tr} \big[{\cal T}_{\alpha \gamma}\big]
(f_{\alpha}-f_{\gamma}), \nonumber \\
\frac{\langle \langle Q^{2}_{\alpha}\rangle \rangle}{t_M}&=&\int_{-\infty}^{\infty} \frac{d\omega}{4\pi}\, (\hbar \omega)^{2} \!\!\sum_{\gamma \neq \alpha =1}^{r} \Big\{ (f_{\alpha} + f_{\gamma} + 2 f_{\alpha} f_{\gamma}) {\rm Tr} \big[{\cal T}_{\alpha \gamma}\big] \nonumber \\
&& + (f_{\alpha}-f_{\gamma})^{2} {\rm Tr} \big[{\cal T}^{2}_{\alpha \gamma}\big]\Big\}.
\end{eqnarray}

\section{Special Case: Two-terminal situation}
In this section we present the CGF and numerical results for exchanged heat for two-terminal case ($r=2$) which we call as the left ($L$) and the right ($R$) heat baths. 
In this case the matrix $\bf{V}$ and $\bf{G}$ introduced in Eq.~(\ref{CGF-2}) are given as (in continuous time)
\begin{equation}  
V = \left( \begin{array}{cc}
                            0 & V^{LR} \\
                             V^{RL} & 0
                         \end{array} \right) \delta(\tau,\tau'), \,\,
G=\left (\begin{array}{cc} 
                              G_{LL} & G_{LR} \\
                              G_{RL} & G_{RR}
                         \end{array} \right),
\end{equation}
and since we are interested in the CGF for heat flowing out of the left lead, only the measurement of the left-lead Hamiltonian is required at two-times and therefore
\begin{equation}
g^{A}=\left (\begin{array}{cc} 
                              g^{A}_{L} & 0 \\
                              0 & 0
                         \end{array} \right).
\end{equation}
For the calculation of entropy-production the second diagonal element of $g^{A}$ will be non-zero. 
Multiplying the matrices the CGF for heat is then given by
\be
{\cal F}(\chi_L)=-\frac{1}{2} {\rm Tr}_{j,\tau} \ln \Big[I-G_{RR}\Sigma^{A}_{L}\Big],
\label{twolead}
\ee
where $\Sigma^{A}_{L}=V^{RL}g_{L}^AV^{LR}$. Similarly the expression for the entropy-production can be obtained. Note that the above expression is similar to the one obtained in Ref.~\onlinecite{fcs-bijay}. However, the meaning of $G^{RR}$ and $\Sigma^{A}_{L}$ are completely different. The long-time limit result can be obtained following the same steps as previous and it is given as
\begin{equation}
{\cal F}(\chi_L)\!=\!-t_M \int_{-\infty}^{\infty} \frac{d\omega}{4\pi} \,\ln \det \Bigl[I - {\cal T}_{RL}[\omega] {\cal K}_{LR}(\chi_L)\Big],
\end{equation}
where the transmission matrix for two-terminal set-up is given as
\begin{equation}
{\cal T}_{RL}[\omega]=G_{RR}^{r} {\tilde \Gamma_{R}} G_{RR}^{a} \Gamma_{L},
\end{equation}
where $\Gamma_{L}=i(\Sigma^{r}_L- \Sigma^{a}_{L})$ with $\Sigma^{r,a}_L= V^{RL} g_{L}^{r,a} V^{LR}$. The expression matches with the result obtained in Ref.~\onlinecite{Huanan} and can be named as the Caroli formula for the ballistic interface. Also in this case it is possible to obtain the two parameter ($\chi_L, \chi_R$) CGF corresponding to the measurement of both left and right lead Hamiltonian simultaneously under the commutative condition of the $\Gamma_L$ and $\tilde{\Gamma}_R$ matrices i.e., $\big[\Gamma_L, \tilde{\Gamma}_R\big]=0$ and can be written as 
\begin{equation}
{\cal F}(\chi_L,\chi_R)\!=\!-t_M \int_{-\infty}^{\infty} \frac{d\omega}{4\pi} \ln \det \Big[I- {\cal T}_{RL}[\omega] {\cal K}_{LR}\big(\chi_{L}-\chi_{R}\big)\Big].
\end{equation}

\subsection{Numerical Results and Discussion}
In this section we present numerical results for the first four cumulants of heat, obtained from Eq.~(\ref{twolead}) by taking derivative of the CGF with respect to the counting field $\chi_{L}$ i.e,  $\langle \langle Q_{L}^{n} \rangle \rangle=\frac{d^{n}{\cal F}(\chi_L)}{d (i\chi_L)^{n}} |_{\chi_L=0}$, as a function of measurement time $t_M$. For performing numerical simulations, we choose two identical semi-infinite $(N_{\alpha} = \infty)$ one-dimensional harmonic chain with nearest-neighbor interactions and are connected via time-dependent coupling. 
In order to obtain the cumulants we need to solve the Dyson equation for $G_{RR}$ obtained
from Eq.~(\ref{Dyson-1}) and the shifted self-energy $\Sigma^{A}_{L}$ in the time-domain
which can be calculated from the surface Green's functions. Since the baths are at their
respective equilibrium temperatures knowing one Green's function is sufficient to obtain the rest. For semi-infinite chain the analytical form for the retarded component of surface Green's function is known and in the frequency domain is written as \cite{Eduardo}
\begin{equation}
g_{ij}^{\alpha,r}[\omega]=-\frac{\lambda}{k} \lambda^{|i-j|}, \quad \alpha=L,R,
\end{equation}
where 
\begin{equation}
\lambda=-\frac{\Omega}{2k} \pm \frac{1}{2k}\sqrt{\Omega^{2}-4k^{2}}= e^{\pm iq},
\end{equation}
where $i,j$ are the site indices, $q$ is the wave vector and $\Omega=(\omega+i\eta)^{2}-2k-k_{0}=-2k \cos q$ is the phonon-dispersion relation for one-dimensional semi-infinite chain. Here $k_0$ is the quadratic on-site potential on each atom. The choice between plus and minus sign is made on satisfying $|\lambda|<1$. The other components can be determined using the relations such as fluctuation-dissipation theorem. For example the lesser component is given by
\begin{equation}
g_{ij}^{\alpha,<}[\omega]=2if_{\alpha} {\rm Im}\big[g_{ij}^{\alpha,r}[\omega]\big],
\end{equation}
and greater component obey the detailed balance condition i.e., $g_{ij}^{\alpha,>}[\omega]=e^{\beta_{\alpha} \hbar \omega}g_{ij}^{\alpha,<}[\omega]$. The inverse Fourier transformation for lesser and greater components are easy to perform as the range of integration in $\omega$ space is confined within the phonon band $k_{0}<\omega^{2}<4k+k_{0}$. 
Therefore the nonequilibrium Green's functions i.e, $G_{RR}^{<,>,r,a}$ can be calculated from the integrals of the equilibrium Green's functions using Eq.~(\ref{Dyson-1}). 
\begin{figure}
\includegraphics[width=\columnwidth]{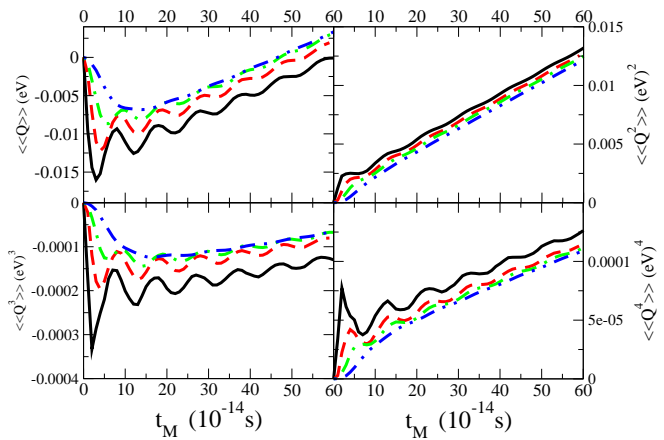}
\caption{(Color online) The cumulants of heat $\langle \langle {Q}_{L}^{n} \rangle
\rangle$ for $n$=1, 2, 3, and 4 for one-dimensional linear chain for two types of
coupling as a function of measurement time. The solid line corresponds to
$V^{LR}(t)=k_{12}\,\theta(t)$. The dash, dash-dotted and dash-dotted-dotted lines
corresponds to $V^{LR}(t)=k_{12} \tanh(\omega_d t)$ with $\omega_d=1.0 [1/t],0.5 [1/t],
0.25 [1/t]$ respectively. The temperatures of the left and the right lead are 310 K and
290 K, respectively.}
\end{figure}
\begin{figure}
\includegraphics[width=\columnwidth]{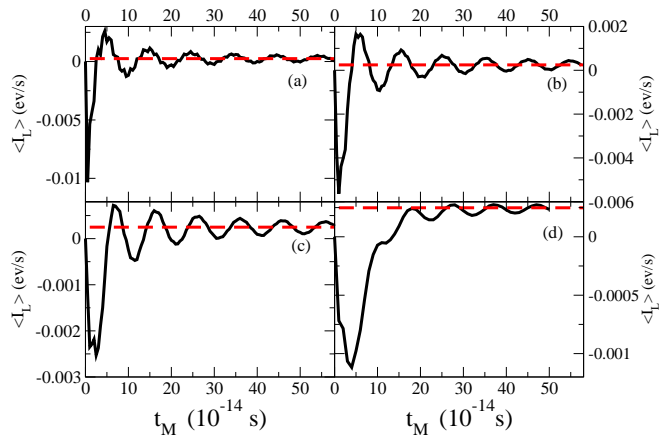}
\caption{(Color online) Plot of current as a function of measurement time $t_M$ for different couplings $V^{LR}(t)$. Figure (a) corresponds to $V^{LR}(t)=\theta(t)\,k_{12}$. Figure (b),(c) and (d) corresponds to $V^{LR}(t)=k_{12} \tanh(\omega_d t)$ with $\omega_d=1.0\,[1/t],0.5\,[1/t], 0.25\,[1/t]$ respectively. The temperatures of the left and the right lead are 310 K and 290 K, respectively. Dotted lines corresponds to the current predicted by Landauer formula.}
\end{figure}

In fig.~3 we show the time-dependent behavior of the first four cumulants of heat $\langle \langle {Q}_L^{n} \rangle \rangle$ for two different forms of the coupling $V^{LR}(t)=\theta(t)\,k_{12}$ and  $V^{LR}(t)=k_{12} \tanh(\omega_d t)$ where $k_{12}$ is the interface force constant. The form of this couplings decides how the device is switched on. In all our numerical calculations, we set the inter-particle and interface coupling $k=k_{12}=1$ eV/(u\AA$^2)$ and the on-site potential $k_0=0.1$ eV/(u\AA$^2)$. These choices of units fix the time scale of our problem which we also use as the unit of time $[t]=10^{-14} s$. The existence of nonzero higher order cumulants $(>2)$ confirms that the probability distribution for the heat is non-Gaussian. We also see that the fluctuations for the cumulants are larger for the sudden switch on case as compared to the slow switch on of the couplings using hyperbolic tangent form. By gradually reducing the driving frequency $\omega_d$ the system evolves to the unique steady state with less and less oscillations. In the long-time limit all cumulants are proportional to the measurement time $t_M$ and the slopes of the cumulants matches with steady-state predictions.

In fig.~4 we plot the current $\langle {\cal I}_{L}(t) \rangle$ by taking derivative of the first cumulant with respect to the measurement time $t_M$. We also obtain the steady-state values of the current using Landuaer formula with unit transmission within the phonon band which are shown with dashed lines. In all cases, at the initial time current is negative, i.e., heat flows into the left lead. Similar situation is also observed for the right lead. The current that goes into the leads is in the form of mechanical energy which is coming from the work that is required to couple both the systems at $t=0$. We also see that at earlier times the amplitude of the current depends on the values of $\omega_d$. Higher driving frequency produce larger transient currents. However at later times the coupling reaches to a constant value $k_{12}$ and the current settles down to the value predicted by Landauer formula.


\subsection{Exchange Fluctuation Theorem (XFT)}
In this section we examine the validity of XFT for different coupling strength $k_{12}$ for the sudden switch on case. This particular fluctuation theorem was first discussed by Jarzynski and W\'ojcik which states that $\langle e^{-\Delta \beta Q_{L}} \rangle_{t_M}=1$ for two weakly connected systems. The relation is true for any transient time $t_M$. We use Eq.~(\ref{twolead}) and calculate the CGF for $\chi_L=i\Delta \beta$. In fig.~5 we show $\langle e^{-\Delta \beta Q_{L}} \rangle$ as a function of $t_M$ for different values of interface coupling strength $k_{12}$ and absolute temperatures $T_L$ and $T_R$ of the baths. For weak coupling ($k_{12}=0.001 k$) the fluctuation symmetry is satisfied and for higher values of the coupling strength the quantity $\langle e^{-\Delta \beta Q_{L}} \rangle -1$ increases. It is important to note that the meaning of weak coupling, in order to satisfy XFT, also depends on the absolute temperatures of the heat baths. This is simply due to the presence of the factor $\Delta \beta Q_L$ in the exponent. Since the cumulants of heat depends on the interface coupling strength (In the long-time limit $\langle \langle Q_L^{n} \rangle \rangle \propto k_{12}^{2}$ in the presence of on-site potential $k_0$), keeping the value of $k_{12}$ constant, if we lower the value of $\Delta \beta$ maintaining the same temperatures difference it is possible to reduce the value of $\langle e^{-\Delta \beta Q_{L}} \rangle -1$, as shown by the dotted line for two values of $k_{12}=1.0, 0.5$ in fig~5. So both the weak-coupling as well as the absolute temperatures of the baths are important to check the validity of XFT. 

Also note that in the long-time limit ($t_M \to \infty$) according to steady state fluctuation theorem (SSFT) \cite{Gallavotti} $\langle e^{-\Delta \beta Q_{L}} \rangle=1$  which is true for arbitrary coupling strength and can be proved trivially from the relation $\langle e^{-\Sigma} \rangle =1$ as in the steady state $Q_L=-Q_R$. In fig.~5, in the long time limit $\langle e^{-\Delta \beta Q_L}\rangle \neq 1$. However this is not a violation of SSFT as the theorem is true only to the leading order of $t_M$. In general the next order correction to the CGF is a constant in $t_M$ \cite{fluc-theorems}
\begin{equation}
{\cal Z}(\chi_L) = e^{-t_M a(\chi_L)} + b(\chi_L) +  {\rm lower \,\, orders \,\,in}\,\, t_M.
\end{equation}
Since $\langle e^{-\Delta \beta Q_{L}} \rangle$ is calculated taking into account the contributions due to all order of $t_M$, it is not equal to one.

\begin{figure}
\includegraphics[width=0.75\columnwidth]{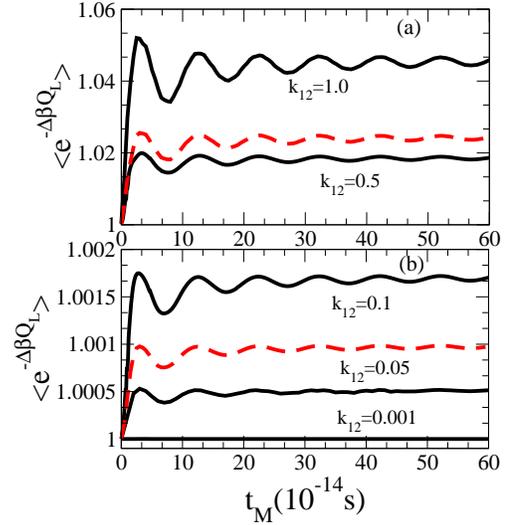}
\caption{(Color online) Plot of $\langle e^{-\Delta \beta Q_{L}} \rangle$ as a function measurement time $t_M$ for different values of interface coupling strength $k_{12}$ and absolute temperatures of the heat baths. The solid lines corresponds to temperatures of the left and the right lead 310 K and 290 K respectively and for dotted lines the temperatures are 510 K and 490 K respectively.}
\end{figure}

\subsection{Effect of finite size of the systems}
In this section we examine the impact of finite size of the systems on the cumulants and current. We only focus on the coupling $V^{LR}(t)=k_{12}\,\theta(t)$. In the case for finite number of oscillators the surface Green's function can also be obtained and is written as
\begin{equation}
g_{\alpha}^{r}[\omega]=-\frac{\lambda}{k} \frac{1-\lambda^{2N_{\alpha}}}{1-\lambda^{2N_{\alpha}+2}}, \quad \alpha=L,R,
\end{equation}
where the meaning of $\lambda$ is the same as before. In the limit $N_{\alpha} \to \infty$ one can recover the surface Green's function for the lead. Unlike the case for the leads $g_{\alpha}^{r}[\omega]$ for finite system is not a smooth function of $\omega$ but rather consists of delta peaks determined by the normal-mode frequencies. Fourier transformation of such functions are difficult to obtain numerically. So we directly evaluate these Green's functions in the time-domain by solving the Heisenberg's equation of motions with fixed boundary conditions. As before knowing one kind of Green's function is enough to determine the rest. So we write down the expression for the greater component as
\begin{equation}
g_{ij}^{\alpha,>}(t)\!=\! -\frac{i}{\hbar} \sum_{k=1}^{N_{\alpha}} \!S^{\alpha}_{ik} \Big [ \frac{\hbar}{2 \omega^{\alpha}_k} (1\!+\!2f^{\alpha}_k)\cos(\omega^{\alpha}_k t) \!-\!\frac{i\hbar}{2} \frac{\sin{\omega^{\alpha}_k t}}{\omega^{\alpha}_k} \Big] S^{\alpha}_{jk},
\end{equation}
where $\big(\omega^{\alpha}_k\big)^{2}=2 k [1-\cos\big(\frac{n \pi}{N+1})] + k_{0}, \,\, n=1,2,\cdots N_{\alpha}$ are the normal mode frequencies and the eigenfunctions are given by
$\epsilon_{j}^{n}= \sqrt{\frac{2}{N+1}} \sin(\frac{n\pi j}{N+1})$. The symmetric matrix $S$ consists of this eigenfunctions which diagonalizes the force constant matrix $K^{\alpha}$ which is a $N_{\alpha} \times N_{\alpha}$ tridiagonal matrix with $2k+k_0$ along the diagonals and $-k$ along the two off-diagonals.
\begin{figure}
\includegraphics[width=\columnwidth]{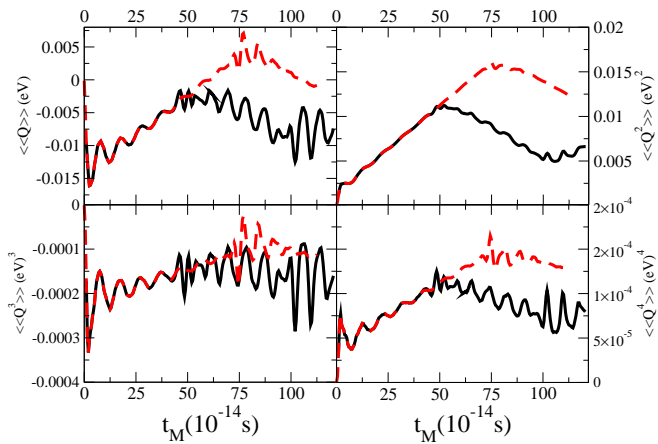}
\caption{(Color online) Plot of the cumulants of heat $\langle \langle {Q}_{L}^{n} \rangle \rangle$ for $n$=1, 2, 3, and 4 for finite harmonic chain. The black and red line corresponds to $N_L=N_R=20$ and $N_L=N_R=30$ respectively. The temperatures of the left and the right lead are 310 K and 290 K, respectively.}
\end{figure}

In fig.~6 we show the results for the cumulants for finite harmonic chain. As we can see that all the cumulants reaches a quasi-steady state values \cite{Di-ventra} with a finite recurrence time $t_r$ which depends on the total length of the system and the velocity of sound waves. After time $t_r$ the phonon modes travelling from $L$ to $R$ reflected from the boundary and then interfere with the incoming waves from $L$ resulting in the cumulants to oscillate rapidly. The importance of on-site potential to achieve quasi-steady state is addressed in Ref.~\onlinecite{Eduardo-finite}. In the limit when the leads are semi-infinite, $t_r \to \infty$ and we observe complete irreversible behavior of the cumulants. The FT's in this case are valid only in the range $0<t<t_r$.

\section{conclusion}
In conclusion, we have extended the study of FCS for heat transport from a lead-junction-lead set-up to a multi-terminal system without the junction part. We found a concise expression for the generalized CGF, valid for arbitrary time $t_M$, under a very general scenario where the coupling between the systems could have arbitrary time-dependent form and also the size of the systems could be finite or infinite. In the stationary state the expression for CGF for heat is obtained. The transient and steady-state fluctuation theorem are explicitly verified.  For two-terminal case the effect on the cumulants and current for two specific form of the coupling are shown. It is also interesting to study the heat-current for other forms of time-dependent coupling as studied in Ref.~\onlinecite{Eduardo} to manipulate the heat flow through the leads and model it to act as a heat-pump. In the long-time limit, invoking time translational invariance we show that the CGF can be expressed in terms of a new transmission function which is useful for the study of interface effects. We also discuss the effect of interface coupling and absolute temperature of the heat baths on XFT which are important for the validity of the theorem. The consequences due to finite size of the systems are also studied. We also note that the theory presented here can be extended to tight binding electrons where it is possible to study both charge and energy transport.

\section*{Acknowledgments}
We are grateful to Juzar Thingna, Meng Lee Leek, Zhang Lifa for insightful discussions. B. K. Agarwalla thanks C. Jarzynski and A. Dhar for stimulating discussion on fluctuation theorems during RRI school on Statistical physics, India. This work is supported in part by a URC research grant R-144-000-257-112 of National University of Singapore.

\appendix

\section{Derivation for Eq.~\eqref{CGF-2}}
In this section we derive Eq.~\eqref{CGF-2} starting from Eq.~\eqref{CGF-1}.
We define a new Green's function ${\bf g^A}= {\bf g^X}- {\bf g}$ and write Eq.~\eqref{CGF-1} as
\be
{\cal F}(\vec{\chi})= -\frac{1}{2} {\rm Tr}_{j,\tau} \ln \big[{\bf I}- {\bf V} {\bf g}- {\bf V} {\bf g^{A}}\big]
\ee
This expression can be further expressed as 
\begin{eqnarray}
{\cal F}(\vec{\chi})&=& -\frac{1}{2} {\rm Tr}_{j,\tau} \ln \big[({\bf g^{-1}}- {\bf V} - {\bf V} {\bf g^{A}} {\bf g^{-1})} {\bf g}\big] \nonumber \\
&=& -\frac{1}{2} {\rm Tr}_{j,\tau} \ln \big[(\bf{G}^{-1} - {\bf V} {\bf g^{A}} {\bf g^{-1}}) {\bf g}\big] \nonumber \\
&=& -\frac{1}{2} {\rm Tr}_{j,\tau} \ln \big[({\bf I}- {\bf V} {\bf g^{A}}{\bf g^{-1}} {\bf G}) {\bf G^{-1}} {\bf g}\big] \nonumber \\
&=& -\frac{1}{2} {\rm Tr}_{j,\tau} \ln \big[{\bf I}- {\bf V} {\bf g^{A}}({\bf I}+ {\bf V} {\bf G})\big]\!-\!\frac{1}{2} {\rm Tr}_{j,\tau} \ln \big[{\bf I}- {\bf V} {\bf g}\big]\nonumber \\
\end{eqnarray}
where we define ${\bf G^{-1}}= {\bf g^{-1}}- {\bf V}$ in contour time which in the integral form is given in Eq.~\eqref{Dyson-1}. We also use the fact that $ {\rm Tr} \ln A = \ln \det A$ and $\det (AB) = \det A \det B$. The term ${\rm Tr} \ln \big[{\bf I}- {\bf V} {\bf g}\big]$ is zero because this corresponds to the sum of vacuum diagrams (counting field independent). Finally using the cyclic property under trace operation we obtain Eq.~\eqref{CGF-2}.

\section{Proof for transient fluctuation theorem}
In this section we proof the transient fluctuation theorem given as $\langle e^{-\Sigma} \rangle =1$ where $\Sigma$ is defined in Eq.~\eqref{sigma}. The important relation that goes into the proof is the
generalized version of the Kubo-Martin-Schwinger (KMS) condition \cite{Kubo} for the
reservoir correlation functions given as
\begin{equation}
g_{\alpha}^{<}(t-\hbar \chi_{\alpha})= g_{\alpha}^{<}(-t+\hbar \chi_{\alpha} + i
\beta_{\alpha} \hbar) = g_{\alpha}^{>}(t-\hbar \chi_{\alpha}-i\beta_{\alpha} \hbar),
\end{equation}
where $\alpha=1,2, \cdots r$. For $\chi_{\alpha}=0$ correlations functions satisfy the standard KMS condition i.e.,
$g_{\alpha}^{<}(t+i\beta_{\alpha}\hbar)=g_{\alpha}^{>}(t)$ which in the frequency domain
reads as the detailed balanced condition i.e., $g_{\alpha}^{>}(\omega)=e^{\beta_{\alpha}
\hbar \omega} g_{\alpha}^{<}(\omega)$. 

To prove the fluctuation theorem we proceed from Eq.~(\ref{CGF-1}) and consider the
$n$-th order term of the log series which in the contour time is given as
\begin{align}
A_{n}= &\int\! d\tau_1 \!\int \!d\tau_2 \!\cdots \!\int\! d\tau_{n} {\rm Tr}_{j}
\Big[V(\tau_1) g^{X}(\tau_1,\!\tau_2) \!\!\cdots\!
V(\tau_{n}) \nonumber \\
& g^{X}(\tau_{n},\!\tau_1)\Big]
\end{align}
where we use the property of the $V$ matrix given as $V(\tau,\tau')=\delta(\tau,\tau'){V}(\tau)$. 

Let us now define a new matrix ${\cal G}_{X}(\tau,\tau')\equiv V(\tau) g^{X}(\tau,\tau')$ and transform from contour time to real time. Then we obtain  
\begin{align}
A_{n}=&\int\! dt_1 \!\!\int \!dt_2 \!\!\cdots \!\int\! dt_{n} \!\!\!\sum_{\sigma_{1} \sigma_{2}
\!\cdots\! \sigma_{n}}\! \!\! {\rm Tr}_{j}\!
\Big[\sigma_1 {\cal G}_{X}^{\sigma_1 \sigma_2} (\tau_1,\!\tau_2)\! \cdots \nonumber \\
& \sigma_n {\cal G}_{X}^{\sigma_n \sigma_1}(\tau_{n},\!\tau_1)\Big]
\end{align}
which can be written in a compact way as
\begin{equation}
A_{n}=\int\! dt_1 \!\!\int \!dt_2 \!\!\cdots \!\int\! dt_{n} 
{\rm Tr}_{j,\sigma}\!
\Big[\bar{{\cal G}}_{X}(t_1,\!t_2)\! \cdots 
\!\bar{{\cal G}}_{X}(t_{n},\!t_1)\Big]
\label{compact-expression}
\end{equation}
where ${\bar{\cal G}_{X}} \equiv \sigma_{z} {\cal G}_{X}$ and $\sigma_z$ is the
third-Pauli matrix. The explicit form of ${\bar{\cal G}_{X}}$ is
\begin{widetext}
\be
{\bar{\cal G}_{X}}(t,t')=\left[\begin{array}{cc}
                                  V(t)\,g^{t}(t\!-\!t') & V(t)\,
g^{<}(t\!-\!t'\!-\!\hbar \chi) \\
                                 -V(t)\,g^{>}(t\!-\!t'\!+\!\hbar \chi)) &
-V(t)\,g^{\bar{t}}(t\!-\!t') 
                               \end{array}
 \right]
\ee
\end{widetext}
Now substituting $\chi_{\alpha}=-i\beta_{\alpha} \,\forall\, \alpha=1,2, \cdots r$ and using KMS boundary condition we
immediately see that the lesser (greater) Green's functions gets interchanged whereas
time-ordered and anti-time ordered Green's functions stays the same. 
Therefore we have 
\be
{\bar{\cal G}}_{\vec{\chi} = -i\vec{\beta}}(t,t')=\left[\begin{array}{cc}
                                 {V}(t)\,g^{t}(t-t') & {V}(t)
\,g^{>}(t-t') \\
                                 -{V}(t)\,g^{<}(t-t') &
-V(t)\,g^{\bar{t}}(t-t') 
                               \end{array}
 \right]
\ee
where ${\vec{\beta}}=(\beta_1,\beta_2 \cdots \beta_r)$.
Now performing an orthogonal Keldysh rotation to this matrix we obtain
\begin{eqnarray}
&&{\breve{\cal G}_{\vec{\chi}=-i\vec{\beta}}}(t,t')=R^{T} {\bar{\cal G}_{\vec{\chi}=-i\vec{\beta}}}(t,t') R \nonumber \\
&&= \left[\begin{array}{cc}
                                 {V}(t)\,g^{a}(t-t') & {V}(t)\,
g^{k}(t-t') \\
                                 0 &
-{V}(t)\,g^{r}(t-t') 
                               \end{array}
 \right]
\end{eqnarray}
where 
\be
R= \frac{1}{\sqrt{2}}\left[\begin{array}{cc}
1 & 1 \\
-1 & 1 
\end{array}
\right]
\ee 
Note that because Keldysh rotation is an orthogonal
transformation the trace in Eq.~\eqref{compact-expression}
remain invariant and we can write
\be
A_{n}=\int\! dt_1 \!\!\int \!dt_2 \!\!\cdots \!\int\! dt_{n} 
{\rm Tr}_{j,\sigma}\!
\Big[\breve{{\cal G}}_{-i\vec{\beta}}(t_1,\!t_2)\! \cdots 
\!\breve{{\cal G}}_{-i\vec{\beta}}(t_{n},\!t_1)\Big],
\ee
Now as the product of the tridiagonal matrices are tridiagonal, then
finally the trace over the branch index $\sigma$ will give
\begin{eqnarray}
A_{n}&=&\int\! dt_1 \!\!\int \!dt_2 \!\!\cdots \!\int\! dt_{n} 
{\rm Tr}_j\Big[{V}(t_1) g^{a}(t_1\!-\!t_2) {V}(t_2) g^{a}(t_2\!-\!t_3)
\cdots \nonumber \\ && {
V}(t_n) g^{a}(t_n\!-\!t_1) + {V}(t_1) g^{r}(t_1\!-\!t_2) {V}(t_2) g^{r}(t_2\!-\!t_3) \cdots \nonumber \\ 
&& {
V}(t_n) g^{r}(t_n\!-\!t_1)\Big].
\end{eqnarray}
This expression is zero because of the causality property of retarded and advanced Green's functions i.e., $g^{r}(t)=0$ for $t<0$ and $g^{a}(t)=0$ for $t>0$. 
This implies $\ln {\cal Z}(\vec{\chi}=-i\vec{\beta})=0$ and thus $\langle e^{-\Sigma}\rangle_{t_M}=1$ for arbitrary $t_M$ and time-dependent
coupling.

\end{document}